\documentclass[aps,onecolumn,showpacs,nofootinbib,showkeys]{revtex4-2}
\usepackage[margin=3.15cm]{geometry}

\RequirePackage[T1]{fontenc}

\usepackage{epsfig,graphicx,amsmath,amssymb,bm}
\RequirePackage{mathptmx}      
\usepackage{mathrsfs}
\usepackage[english]{babel}
\usepackage{amssymb}
\RequirePackage{color}
\usepackage{changes}
\usepackage{slashed}
\usepackage{multirow}
\usepackage{scalerel}
\usepackage{tikz-feynman}
\usepackage{textcomp}
\usepackage{subcaption}
\usepackage[english]{babel}
\usepackage{float}
\RequirePackage{hyperref}
\hypersetup{
    linktocpage,
    colorlinks,
    citecolor=blue,
    filecolor=black,
    linkcolor=blue,
    urlcolor=blue,
}
\usepackage{nccmath}

\begin{document}

\title{The decay $\eta' \to \pi\rho$ in the chiral NJL model}


\author{M. K. Volkov$^{1}$}\email{volkov@theor.jinr.ru}
\author{A. A. Pivovarov$^{1}$}\email{tex$\_$k@mail.ru}
\author{K. Nurlan$^{1,2,3}$}\email{nurlan@theor.jinr.ru}

\affiliation{$^1$ BLTP, Joint Institute for Nuclear Research, Dubna, 141980 Russia \\
                $^2$ Institute of Nuclear Physics, Almaty, 050032, Kazakhstan \\
                $^3$ L.N. Gumilyov Eurasian National University, Nur-Sultan, 01008, Kazakhstan}   


\begin{abstract}
The branching fraction of the decay $\eta' \to \pi\rho$ is calculated in the chiral quark NJL model. The decay exists due to the mass difference between $u$- and $d$-quarks leading to $\pi^{0}-\eta^{'}$ mixing. Two different approaches are applied: the approximate calculation of this process taking into account the transition $\pi^{0}-\eta^{'}$ explicitly without dioganalisation and a more precise approach consisting in the diagonalisation of the singlet and octet states leading to the physical fields $\pi^{0}$, $\eta$ and $\eta^{'}$. The obtained results are in satisfactory agreement with current experimental data.


\end{abstract}


\maketitle



Recently, the collaboration BESIII reported new experimental data for the decays $\eta' \to \pi^+\pi^- \pi^0$ and $\eta' \to \pi^0\pi^0 \pi^0$ \cite{BESIII:2016tdb}. These experimental measurements allowed one to estimate the partial width of the decay $\eta' \to \pi^{\pm} \rho^{\mp}$ with the result $Br(\eta' \to \pi^{\pm} \rho^{\mp}) = (7.44 \pm 0.60)\times 10^{-4}$. This process is a dominant intermediate channel of the decay $\eta' \to \pi^+\pi^- \pi^0$\cite{BESIII:2016tdb}. The theoretical description of the decay $\eta' \to \pi^+\pi^- \pi^0$ was presented in the work \cite{Borasoy:2006uv}.  

The decay $\eta' \to \pi \rho$ is interesting for the following reasons. It exists due to the mass difference between the $u$- and $d$-quarks. As a result, a possibility of transition between the fields $\pi$ and $\eta'$ appears. In present time, there is a significant difference in the estimations of the value of this transition (see Table \ref{tab_1}). The calculation of the decay $\eta' \to \pi \rho$ is a good possibility to verify these estimations. Indeed, the width of this decay depends on two parameters: the constant $g_{\rho} = 6$ of the decay $\rho \to \pi \pi$ \cite{Volkov:2021fmi, Volkov:2022jfr} and the value of the $\pi^{0}-\eta'$ transition. For this reason, in the present work, the main attention is paid to the consideration of this decay from different points of view.

In the present paper, the calculation of the decay $\eta' \to \pi \rho$ is carried out in the framework of the Nambu--Jona-Lasinio (NJL) model \cite{Volkov:1986zb, Ebert:1985kz, Vogl:1991qt, Ebert:1994mf, Volkov:2005kw, Volkov:2022jfr}. The difference between the $u$- and $d$-quark masses determining the value of this decay can be found by using the experimental width of the non-standard decay $\omega \to \pi \pi$. The corresponding calculation was described in the review \cite{Volkov:1986zb} and then with the new corrected parameters in the recent review \cite{Volkov:2022jfr}. As a result, the value 4.2 MeV has been obtained for the quark mass difference.

The found quark mass difference allows one to calculate the values of the transitions $\pi^{0}-\eta$ and $\pi^{0}-\eta^{'}$. These transitions also make such a decay as $\rho \to \pi \eta$ possible. This process was described in the paper \cite{Volkov:2012be} and also in the paper \cite{Paver:2010mz}. More precise estimations for this process will be given in this work. Unfortunately, at the present time, the experimental estimation of this process gives only the upper limit. That is why, the main purpose of the present paper is to describe the decay $\eta' \to \pi \rho$.

\begin{figure*}[t]
 \centering
 \centering
 \begin{subfigure}{0.5\textwidth}
  \centering
   \begin{tikzpicture}
    \begin{feynman}
      \vertex (a) {\(\pi^0\)};
      \vertex [dot, right=1.3cm of a] (b){};
      \vertex [dot, right=1.2cm of b] (c){};
      \vertex [right=1.3cm of c] (d){\(\eta(\eta')\)};
      \diagram* {
         (a) -- [] (b),
         (b) -- [fermion, inner sep=1pt, half left] (c),
         (c) -- [fermion, inner sep=1pt, half left] (b),         
         (c) -- [] (d),
      };
     \end{feynman}
    \end{tikzpicture}
  \end{subfigure}%
 \caption{Diagram describing $\pi^0 \to \eta(\eta')$ transitions in the NJL model.}
 \label{diagram1}
\end{figure*}
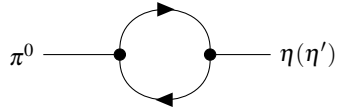

For a description of the transitions $\pi^{0}-\eta$ and $\pi^{0}-\eta'$ we use the quark one-loop approximation with the parameters taken from the NJL model \cite{Volkov:2022jfr} (Fig. \ref{diagram1}). The description of these transitions is carried out by using the parts of the Lagrangian presented bellow containing the interactions of the $\pi$, $\eta$, and $\eta'$ mesons with quarks. Also, we present the Lagrangian of the interaction of the vector $\rho$-meson with quarks which is necessary for the description of the decay $\eta' \to \pi \rho$.

    \begin{eqnarray}
    \label{L1}
    && \Delta{\mathcal L} = \bar{q}\biggl[i g_{\pi}\gamma^5 \left( \lambda_0^{\pi}\pi + \lambda_{+}^{\pi}\pi + \lambda_{-}^{\pi}\pi^{-} \right)
    + \frac{g_\rho}{2}\gamma^{\mu} \left(\lambda_{-}^{\rho}{\rho}^{-}_{\mu}+\lambda_{+}^{\rho}{\rho}^{+}_{\mu}\right) 
    + i \sin(\bar{\theta}) g_{\eta_{u}} \gamma^{5}  \lambda_{u} \eta
    \nonumber
     \\ && \qquad  
      + i \cos(\bar{\theta}) g_{\eta_{s}} \gamma^{5}  \lambda_{s} \eta
      + i \cos(\bar{\theta}) g_{\eta_{u}} \gamma^{5}  \lambda_{u} \eta' - i \sin(\bar{\theta}) g_{\eta_{s}} \gamma^{5}  \lambda_{s} \eta'
     \biggr]q,
    \end{eqnarray} 
    where $q$ and $\bar{q}$ are fields of u, d and s quarks with constituent masses $m_{u} = 270$~MeV, $m_{d} = 274.2$~MeV and $m_ {s} = 420$~MeV; $\lambda$ matrices are linear combinations of the Gell-Mann matrices \cite{Volkov:2022jfr}. $\bar{\theta}=\theta_0 - \theta$, where $\theta=-19^{\circ}$ is the deviation from the ideal mixing angle $\theta_0 = 35.3^{\circ}$ for the $\eta$ and $\eta'$  mesons \cite{Volkov:2022jfr}.

      The renormalization constants for the meson fields take the form
        \begin{eqnarray}
    \label{Couplings}
    	g_{\pi}= \sqrt{\frac{Z_{\pi}}{4I_{2}(u)}}, \quad g_{\rho} = \sqrt{\frac{3}{2I_{2}(u)}}, \quad g_{\eta_{u}}= \sqrt{\frac{Z_{\eta_u}}{4I_{2}(u)}}, \quad g_{\eta_s}= \sqrt{\frac{Z_{\eta_s}}{4I_{2}(s)}}. 
    \end{eqnarray}
    where $Z_{\pi}$, $Z_{\eta_u}$ and $Z_{\eta_s}$ are additional renormalization constants arising in the transitions $\pi-a_{1}$ and $\eta(\eta ')-f_1(1285)(f_1(1420))$ \cite{Volkov:1986zb,Volkov:2022jfr}.
    
  The integrals appearing in the coupling constants have the form
    \begin{eqnarray}
    	I_{n}(q) =
    	-i\frac{N_{c}}{(2\pi)^{4}}\int\frac{\Theta(\Lambda_4^2 + k^{2})}{(m_{q}^{2} - k^2)^{n}}
    	\mathrm{d}^{4}k,
    \end{eqnarray}
where $\Lambda_4 = 1.65$~GeV is the ultraviolet cutoff parameter ~\cite{Volkov:2022jfr}.
    
Using the Lagrangian in (\ref{L1}) for the constant describing the $\pi-\eta$ transition, we obtain the following expression:
\begin{eqnarray}
     \epsilon_{\pi\eta} = \frac{3g_{\pi}^{2} \sin{(\bar{\theta})}}{M_{\eta}^{2} - M_{\pi}^{2}} \left[\int \frac{Tr\left\{\gamma^{5}\left(\hat{k} + \frac{\hat{p}}{2} + m_{u}\right)\gamma^{5}\left(\hat{k} - \frac{\hat{p}}{2} + m_{u}\right)\right\}}{\left[\left(k + \frac{p}{2}\right)^{2} - m_{u}^{2}\right]\left[\left(k - \frac{p}{2}\right)^{2} - m_{u}^{2}\right]} \frac{d^{4}k}{\left(2\pi\right)^{4}} - \int \frac{Tr\left\{\gamma^{5}\left(\hat{k} + \frac{\hat{p}}{2} + m_{d}\right)\gamma^{5}\left(\hat{k} - \frac{\hat{p}}{2} + m_{d}\right)\right\}}{\left[\left(k + \frac{p}{2}\right)^{2} - m_{d}^{2}\right]\left[\left(k - \frac{p}{2}\right)^{2} - m_{d}^{2}\right]} \frac{d^{4}k}{\left(2\pi\right)^{4}}\right].& \nonumber\\
\end{eqnarray}

We use the standard procedure for the NJL model and expand these integrals in a series in powers of external momentum and keep only the divergent parts. Then the transition takes the form:
\begin{eqnarray}
     \epsilon_{\pi\eta} = \frac{g_{\pi}^{2} \sin{(\bar{\theta})}}{M_{\eta}^{2} - M_{\pi}^{2}}\left[4 I_{1}(u) + 2 M_{\eta}^{2} I_{2}(u) - \left(4 I_{1}(d) + 2 M_{\eta}^{2} I_{2}(d)\right)\right].
\end{eqnarray}

Having carried out minor transformations of this expression and excluding converging terms, we can obtain for this transition:
\begin{eqnarray}
     \epsilon_{\pi\eta} = 4 g_{\pi}^{2} \sin{(\bar{\theta})} \left(m_{d}^{2} - m_{u}^{2}\right) I_{2}(u) \frac{1}{M_{\eta}^{2} - M_{\pi}^{2}} \approx (0.93 \pm 0.15) \cdot 10^{-2}.
\end{eqnarray}

For the $\pi-\eta'$ transition, we similarly obtain
\begin{eqnarray}
     \epsilon_{\pi\eta'} = 4 g_{\pi}^{2} \cos{(\bar{\theta})} \left(m_{d}^{2} - m_{u}^{2}\right) I_{2}(u) \frac{1}{M_{\eta'}^{2} - M_{\pi}^{2}} \approx (2.07 \pm 0.35) \cdot 10^{-3}.
\end{eqnarray}

The inaccuracy of the version of the NJL model used in this paper is estimated at $\sim$ 17\% \cite{Volkov:2022jfr}.

The values $\epsilon_{\pi\eta}$ and $\epsilon_{\pi\eta'}$ can be considered as rotation angles of mixing of the states $\pi^{0}-\eta$ and $\pi^{0}-\eta^{'}$.

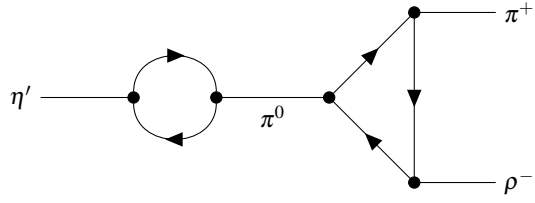
\begin{figure*}[t]
 \centering
 \centering
 \begin{subfigure}{0.5\textwidth}
  \centering
   \begin{tikzpicture}
    \begin{feynman}
      \vertex (a) {\(\eta' \)};
      \vertex [dot, right=1.5cm of a] (b){};
      \vertex [dot, right=1.1cm of b] (c){};
      \vertex [dot, right=1.5cm of c] (d){};
      \vertex [dot, above right=1.6cm of d] (e) {};      
      \vertex [dot, below right=1.6cm of d] (f) {};
      \vertex [right=1.4cm of e] (g) {\(\pi^{+} \)};
      \vertex [right=1.4cm of f] (h) {\(\rho^{-} \)};       
      \diagram* {
         (a) -- [] (b),
         (b) -- [fermion, inner sep=1pt, half left] (c),
         (c) -- [fermion, inner sep=1pt, half left] (b),         
         (c) -- [edge label'=\(\pi^0 \)] (d),
         (d) -- [fermion] (e),  
         (e) -- [fermion] (f),    
         (f) -- [fermion] (d),  
         (e) -- (g),          
         (f) -- (h),         
      };
     \end{feynman}
    \end{tikzpicture}
  \end{subfigure}%
 \caption{Diagram for $\eta' \to \pi^{+} \rho^{-}$ decay in the NJL model.}
 \label{diagram2}
\end{figure*}%

Using the obtained transition value $\pi^{0}-\eta$, for the decay $\rho \to \pi \eta$ we obtain the amplitude of the following form:
\begin{eqnarray}
\label{amplitude1}
    \mathcal{M}(\rho^\pm \to \pi^\pm \eta) = g_{\rho} \epsilon_{\pi\eta} e_{\mu}(p_{\rho}) (p_{\eta} - p_{\pi})^{\mu}.
\end{eqnarray}

Then the branching fraction of this process is:

\begin{eqnarray}
     Br(\rho^\pm \to \pi^{\pm} \eta) \approx (6.29 \pm 1.07) \times 10^{-6}.
\end{eqnarray}

The experiment gives only the upper limit \cite{PDG}:
\begin{eqnarray}
     Br(\rho^{\pm} \to \pi^{\pm} \eta)_{exp} < 6 \cdot 10^{-3}.
\end{eqnarray}

The diagram describing the decay $\eta' \to \pi^{+} \rho^{-}$ is presented in Figure \ref{diagram2}. For this decay, we obtain the following expression:
\begin{eqnarray}
\label{amplitude2}
    \mathcal{M}(\eta' \to \pi^{+} \rho^{-}) = 2g_{\rho} \epsilon_{\pi\eta^{'}} e^{*}_{\mu}(p_{\rho}) p_{\pi}^{\mu}.
\end{eqnarray}

For the $\eta' \to \pi^{-} \rho^{+}$ process, the amplitude looks similar. The resulting branching fraction will be the sum of the partial widths of these processes:
\begin{eqnarray}
     Br(\eta' \to \pi^{\pm} \rho^{\mp}) \approx (5.3 \pm 0.9) \times 10^{-4}.
\end{eqnarray}

The experimental data \cite{PDG}:
\begin{eqnarray}
     Br(\eta' \to \pi^{\pm} \rho^{\mp})_{exp} = (7.4 \pm 2.3) \times 10^{-4}.
\end{eqnarray}

As we can see, our result is in satisfactory agreement with the experiment taking into account the accuracies of measurement and model.

It is interesting to note that in the case of the $\eta' \to \pi \rho$ decay, there is an experimental value of the branching fractions, while for the $\rho \to \pi \eta$ decay, there is currently only the upper bound.

The above approach took into account the mixing between $\eta-\pi^{0}$ and $\eta^{'}-\pi^{0}$ mesons. The states $\eta$ and $\eta^{'}$ were already considered to be diagonalized after mixing due to the 't Hooft interaction \cite{Volkov:2005kw, Vogl:1991qt}. A more precise approach would be to consider mixing of all three states at the same time. For these purposes, the singlet state $\phi_{0}$ and the octet states $\phi_{3}$ and $\phi_{8}$ can be used. The quark-meson Lagrangian containing these states takes the following form:

\begin{eqnarray}
    \Delta{\mathcal L} = i \bar{q} \gamma^{5} \left[\lambda_{0}\phi_{0} + \lambda_{3}\phi_{3} + \lambda_{8}\phi_{8}\right]q,
\end{eqnarray}
where
\begin{eqnarray}
    \lambda_{0} = \sqrt{\frac{2}{3}}\left(
    \begin{tabular}{ccc}
        1 & 0 & 0 \\
        0 & 1 & 0 \\
        0 & 0 & 1
    \end{tabular}
    \right),
    \quad
    \lambda_{3} = \left(
    \begin{tabular}{ccc}
        1 &  0 & 0 \\
        0 & -1 & 0 \\
        0 &  0 & 0
    \end{tabular}
    \right),
    \quad
    \lambda_{8} = \frac{1}{\sqrt{3}} \left(
    \begin{tabular}{ccc}
        1 & 0 &  0 \\
        0 & 1 &  0 \\
        0 & 0 & -2
    \end{tabular}
    \right).
\end{eqnarray}

The fields $\phi_{0}$, $\phi_{3}$ and $\phi_{8}$ can be mixed. In this case, for a correct description of the mixing of the $\phi_{0}$ and $\phi_{8}$ states, it is necessary to take into account the 't Hooft interaction in the initial four-quark Lagrangian. This results in a deviation from the ideal mixing angle by $\theta = -19^{\circ}$. Taking into account the mass difference between the $u$- and $d$-quarks leads to the mixing of the $\phi_{0}-\phi_{3}$ and $\phi_{8}-\phi_{3}$ fields. Since this difference is small, the angle of rotation in the $\phi_{0}-\phi_{3}$ and $\phi_{8}-\phi_{3}$ planes will also be small, and therefore, the field transformation can be presented in the following form (as it was done, for example, in \cite{Osipov:2022bxg,Kroll:2005sd}):

\begin{eqnarray}
    \phi_{3} & = & \pi^{0} - \epsilon_{\pi\eta} \eta - \epsilon_{\pi\eta'} \eta', \nonumber\\
    \phi_{8} & = & \left(\epsilon_{\pi\eta} \cos{\theta} + \epsilon_{\pi\eta'} \sin{\theta}\right) \pi^{0} + \cos{\theta} \eta + \sin{\theta} \eta', \nonumber\\
    \phi_{0} & = & \left(\epsilon_{\pi\eta'} \cos{\theta} - \epsilon_{\pi\eta} \sin{\theta}\right) \pi^{0} - \sin{\theta} \eta + \cos{\theta} \eta',
\end{eqnarray}
where $\epsilon_{\pi\eta}$ and $\epsilon_{\pi\eta'}$ are the rotation angles in the planes $\phi_{0}-\phi_{3}$ and $\phi_{8} -\phi_{3}$. Then for these angles one can obtain the expressions
\begin{eqnarray}
    \epsilon_{\pi\eta} = \frac{M_{03} \sin{\theta} - M_{38} \cos{\theta}}{M_{\eta}^{2} - M_{\pi}^{2}}, \quad \epsilon_{\pi\eta'} = - \frac{M_{03} \cos{\theta} + M_{38} \sin{\theta}}{M_{\eta'}^{2} - M_{\pi}^{2}},
\end{eqnarray}
where $M_{03}$ and $M_{38}$ are the transition amplitudes between the corresponding states. Taking into account the diagram shown in Figure \ref{diagram1}, the following expressions are obtained for these amplitudes:
\begin{eqnarray}
    M_{03} = M_{38} = 4 g_{\pi}^{2} \left(m_{d}^{2} - m_{u}^{2}\right) I_{2}(u).
\end{eqnarray}

Then the desired angles are $\epsilon_{\pi\eta} \approx (1.45 \pm 0.25) \times 10^{-2}$, $\epsilon_{\pi\eta'} \approx (2.21 \pm 0.38) \times 10^{-3}$. Substituting these values into the amplitudes (\ref{amplitude1}) and (\ref{amplitude2}), one can obtain for the decays $\rho^\pm \to \pi^\pm \eta$ and $\eta' \to \pi ^{\pm} \rho^{\mp}$ the following results:
\begin{eqnarray}
     Br(\rho^{\pm} \to \pi^{\pm}\eta ) \approx (1.54 \pm 0.26) \times 10^{-5}, \\
     Br(\eta' \to \pi^{\pm} \rho^{\mp}) \approx (6.00 \pm 1.01) \times 10^{-4}.
\end{eqnarray}

As we can see, the obtained results still do not contradict the experimental data.
  
\begin{table}[h!]
\begin{center}
\begin{tabular}{ccccc}
\hline
 & NJL & Paver and Raizuddin \cite{Paver:2010mz,Paver:2011md} & Osipov \cite{Osipov:2022bxg}  & Kroll \cite{Kroll:2005sd} \\
\hline
$\epsilon_{\pi\eta} \times 10^2$		& $0.93 \pm 0.15 (1.45 \pm 0.25)$    	& $1.34$            &$1.3 \pm 0.1$       & $1.7 \pm 0.2$  \\
$\epsilon_{\pi\eta'} \times 10^3$	& $2.07 \pm 0.35(2.21 \pm 0.38)$	  & $3\pm 1$           &$3.9 \pm 0.3$       &   $4.0 \pm 1.0$ \\
\hline
\end{tabular}
\end{center}
\caption{Comparison of the values of the rotation angles $\epsilon_{\pi\eta}$ and $\epsilon_{\pi\eta'}$. For the case of the NJL model, two results obtained by different methods are presented.}
\label{tab_1}
\end{table}  

The obtained values for the rotation angles $\epsilon_{\pi\eta}$ and $\epsilon_{\pi\eta'}$ can be compared with the results of other authors \cite{Paver:2010mz, Osipov:2022bxg, Paver:2011md, Kroll:2005sd} (see Table \ref{tab_1}). 

Our results have been obtained without introducing additional arbitrary parameters and are in satisfactory agreement with the experimental data for the $\eta' \to \pi^{\pm} \rho^{\mp}$ process. The existing discrepancies with other theoretical works for the values $\epsilon_{\pi\eta}$ and $\epsilon_{\pi\eta'}$ can be explained by the use of different models.

\subsection*{Acknowledgments}
The authors are grateful to A.B. Arbuzov and A.A. Osipov for useful discussions.

\end{document}